\documentstyle[12pt]{article}
\title{Technicolor  leptoquarks and the excess of NC and CC events with 
high-$Q^2$ at HERA}
\author{ Zhenjun Xiao
\thanks{Email: xiaozj@v2.rl.ac.uk} \\  
{\small Rutherford Appleton Laboratory,
Chilton, OXON, OX11 0QX, UK\thanks{Mailing address before 17 Oct. 1997}}\\
{\small Department of Physics, Henan Normal University,
Xinxiang, 453002 P.R.China.\thanks{Permanent address} }\\ }
\date{\today}

\begin{document}
\maketitle
\begin{picture}(0,0)(0,0)
\put(310,260){{\large RAL-TR-97-043}}
\put(310,240){{\large HEP-PH/9709xxx}}
\end{picture}
\begin{abstract}
In this paper we pursue a consistent leptoquark interpretation for 
HERA anomalies in the framework of Technicolor.
We find that: (a) one  F=0 scalar Technicolor leptoquark $P_3^{'}$ with a 
mass of 200 GeV can provide the required contributions to account for 
the excess of both neutral and charged current events with high-$Q^2$
 at HERA; 
(b) the current data still allow the coexistence of  $ P_3^{'}$ with 
$m(P_3^{'})= 200 GeV$ and 
 $P_3^0$ with $m(P_3^0)=225 GeV$,  they could contribute 
effectively to $e^+p$ collision  process and may be 
responsible for the apparent
splitting of average mass of H1 and ZEUS NC events with high-$Q^2$. 
\end{abstract}
\vspace{1cm}

\newcommand{\beq}{\begin{eqnarray}}
\newcommand{\eeq}{\end{eqnarray}}
\newcommand{\qs}{$Q^2$}
\newcommand{\pbm}{$pb^(-1)$}
\newcommand{\empp}{e^{\pm}p}
\newcommand{\emp}{e^-p}
\newcommand{\epp}{e^+p}

\newcommand{\pube}{P_{\overline{U}E}}
\newcommand{\pueb}{P_{U\overline{E}}}
\newcommand{\pubn}{P_{\overline{U}N}}
\newcommand{\punb}{P_{U\overline{N}}}

\newcommand{\pdbe}{P_{\overline{D}E}}
\newcommand{\pdeb}{P_{D\overline{E}}}
\newcommand{\pdbn}{P_{\overline{D}N}}
\newcommand{\pdnb}{P_{D\overline{N}}}
\newcommand{\pto}{P_3^0}
\newcommand{\ptp}{P_3^{'}}
\newcommand{\epd}{e^+d}
\newcommand{\nubu}{\overline{\nu}u}
\newcommand{\breda}{Br(\ptp \to e^+d)}
\newcommand{\brnua}{Br(\ptp \to \overline{\nu}u)}
\newcommand{\bredb}{Br(\pto \to e^+d)}
\newcommand{\brnub}{Br(\pto \to \overline{\nu}u)}
\newcommand{\slq}{\sigma_{NC}} 
\newcommand{\scc}{\sigma_{CC}}

\newcommand{\etal}{\it et\ \  al.,}

\newpage

\section*{1. Introduction}

In this February, the two HERA collaborations H1 \cite{h1} 
and ZEUS \cite{zeus1} reported the excess
of neutral current (NC) events with high-\qs ($Q^2 > 15000 GeV^2$) 
compared with the Standard Model(SM) expectations, based
on their 1994-96 ${e^+p}$ data.
This observation has triggered extensive investigations about the possible 
new mechanisms beyond the SM responsible for this excess \cite{fram}. 
Among many 
possible solutions, the most favored one  is the resonant production of new
bosonic particle with a mass of about $200 GeV$, namely the 
{\it Leptoquark} (LQ)\cite{lq1,rizzo} predicted by Technicolor (TC) and 
other new physics theories \cite{rizzo} or the {\it Squarks} ( $\tilde{t}$ 
and/or  $\tilde{c}$) in the Minimal Supersymmetric 
Standard Model(MSSM) with $R_p$ violating interactions 
\cite{rpv,dreiner}. But the newest  experimental results
seem to disfavor  the leptoquark interpretation based on the 
popular leptoquark scenario \cite{buchmuller87}.

Firstly, the CDF \cite{cdf210} and D0 \cite{d0225} 
collaborations  reported  their 
new lower mass bounds  on scalar leptoquarks ( and 
$R_p$ violating $\tilde{q}$'s ) very recently, 
based on their negative searches for the pair production of first
generation scalar leptoquarks using the full RUN I data set and the new NLO 
theoretical cross section \cite{kramer97}. Assuming 
$\beta=Br(S \to eq)=1$, the CDF and D0  limit is  $M_S \geq 213\ \ 
GeV$ \cite{cdf210} and $M_S \geq 225\ \ GeV$ \cite{d0225} respectively, 
and the combined limit is $M_S \geq 240 GeV$ \cite{kramer972}.  
The corresponding lower mass bounds 
on vector leptoquarks as well as squarks are in general significantly higher 
than that for scalar leptoquarks because the corresponding production cross 
sections for these particles are much more larger than that for scalar 
leptoquarks. 

Secondly, H1 and ZEUS Collaborations \cite{straub,bruel} 
reported more high-$Q^2$ neutral current and charge current (CC)  
events at the 1997 Lepton-Photon conference according to  their new data 
obtained through June 1997. With a combined luminosity of  $57.2\ \ Pb^{-1}$
the former reported excess
of high-$Q^2$ NC events is  supported by the new data, 
and a clear tendency for the CC data  to be above the SM deep inelastic 
scattering (DIS) expectations at large $Q^2$ was found by H1 and ZEUS 
collaborations \cite{straub,bruel}.  
For $Q^2 > 10^4 GeV^2$, they observed 28 CC events where $17.7 \pm 4.3$ 
are expected.  We now have to explain the excess for both NC and CC events 
simultaneously.

Moreover, the invariant mass distributions of the high-$Q^2$ NC events of H1 
and ZEUS are  rather different. For the 1994-96 $\epp$ data, the 7 H1 events 
appear clustered around $M\approx 200 GeV$, while the 5 ZEUS events clustered at 
$M \approx 220 GeV$. Recent studies \cite{bassler,straub} 
showed that this splitting can not be 
accounted for either by initial state radiation (ISR) or by detector effects, 
and it is unlikely that excesses observed by H1 and ZEUS could be caused by
the production and decay of a single narrow resonance \cite{straub}.

Although a scalar leptoquark with a mass of about 
200 GeV is  still  allowed by the Tevatron  data if it can decay 
to other channels, but the F=0 scalar leptoquarks ${\cal R}$ and {\cal 
$\tilde{R}$} in the Buchmuller, R\"ukle and Wyler (BRW) leptoquark 
scenario \cite{buchmuller87} could not
decay to both $e^+d$ and $\nubu$ channels and therefore can not contribute to
both NC and CC processes at HERA simultaneously. In order to provide a 
consistent interpretation  for HERA anomalies, one has to study something 
new beyond the classic BRW leptoquark scenario 
\cite{altarelli,babu5414,rizzo97}.

In this paper, we will  consider: (a) mixed states of color-triplet 
pseudo-Goldstone bosons, (b) more than one TC 
leptoquarks contribute effectively, 
in the framework of Technicolor \cite{farhi,eichten86}. 
We will  calculate the contributions to both NC and CC 
$\empp$ collision processes at HERA from the F=0 scalar Technicolor (TC) 
leptoquark $\ptp$ and $\pto$, the  mixed states of  color-triplet 
pseudo-Goldstone bosons $\punb$ and $\pdeb$.
We find that: (a) one $\ptp$ with mass of 200 GeV can 
provide the required extra contributions to account for the HERA anomalies. 
For $Q^2_{min}=15000 GeV^2$, we have $\sigma_{LQ}^{NC}=0.216\; pb$ and 
$\sigma_{LQ}^{CC}=0.306\; pb$ assuming $m(\ptp)=200 GeV$, $F_{2L}=0.02$ 
and $\beta_{NC}=0.7$;   
(b) the current data still allow the 
coexistence of a $\ptp$ with $m(\ptp)=200 GeV$ and a $\pto$ with 
$m(\pto)=225 GeV$, they could contribute the required extra cross sections 
 to both NC and CC 
processes and may be responsible for the apparent splitting of average 
mass of H1 and ZEUS NC events with very high $Q^2$. 
We also estimated the possible 
contributions from other heavier F=0 scalar TC leptoquarks.

This paper is organized as follows. in Sec.2 we briefly review the masses and
the couplings of TC leptoquarks and the relevant experimental constraints.  In
Sec.3 we calculate the contributions to the $\empp$ collision processes
 from the 
F=0 scalar TC leptoquarks and present the numerical results. Section 4 
contains the conclusion and discussions.

\section*{2. Technicolor leptoquarks, masses and couplings}

As is  well known, the color triplet pseudo-Goldstone bosons (i.e. the 
leptoquarks in TC models) appeared in  almost all non-minimal
TC models which include one generation or more technifermions, 
such as the one generation TC model \cite{farhi},  
one-family $SU(2)_{TC}$ model \cite{appel1} and the  
Postmodern TC model \cite{appel1}, etc. For definiteness, we calculate the 
possible contributions to $\empp$ scattering process from 
F=0 scalar TC leptoquarks 
as described  in the often-discussed 
Farhi-Susskind one generation technicolor model (OGTM) and 
follow the nomenclature defined in ref.\cite{eichten86}. 
Although this model is not rich enough to describe the real world 
\footnote{for the S parameter problem and the current status of 
Technicolor theory, see a recent review \cite{lane96}.}, 
it does provide a typical description for the production and 
decays of such leptoquarks 
which would have to be present in any realistic TC models. 
For our studies in this paper, what we care most are 
the masses of TC leptoquarks and their effective Yukawa couplings to 
lepton-quark pairs, as well as the mixing patterns.

Under the gauge group $ 
SU(N)_{TC}\otimes SU(3)_C \otimes SU(2)_L \otimes U(1)_Y$, the technifermions
transform as \cite{eichten86} 
\beq
\begin{array}{ccl}
{Q_L} =
\left( \begin{array}{c}U_L \\ D_L \end{array}
\right) &\sim& (N,3,2,1/3),\\
{U_R} \sim (N,3,1,4/3),& &
{D_R} \sim (N,3,1,-2/3),\\
{L_L} =\left( \begin{array}{c}N_{L} \\ E_L \end{array}
\right) &\sim& (N,1,2,-1),\\
{E_R} \sim (N,1,1,-2), &&  {N_R} \sim (N,1,1,0), 
\end{array}
\eeq
where the techniquarks and technileptons have the same charges as those 
ordinary quarks and leptons. When the technifermion condensate 
$<\overline{T}T>\ne 0$ is formed, the global flavor chiral symmetry
$ SU(8)_L\otimes SU(8)_R$ is broken down to $SU(8)_{L+R}$.  
Consequently, 63 pseudo-Goldstone bosons(PGB's) would be produced. 
Corresponding to each of these $^1S_0$ pseudoscalars is a hyperfine partner 
$^3S_1$ technirho. 
Among all these PGB's and technirhos, the color-triplet 
$Q\overline{L}$ and $\overline{Q}L$ bound states 
are what we are most interested in for 
HERA experiments. Since only those F=0 TC leptoquarks may have a sizable 
contribution to the $\epp$ collision process at HERA we will 
not consider other PGB's and technirhos in this paper.

In Table 1 we classify the TC leptoquarks according to their $SU(3)_c$ and 
$SU(2)_v$ quantum numbers. Among all color-triplets the technirhos $\rho_3$ 
are  the usual $F=0$ vector leptoquarks, and they form one isotriplet 
$(\rho^1_3, \rho^0_3, \rho^{-1}_3)$ and isosinglet $\rho^{'}_3$. These vector 
leptoquarks may acquire large masses from the strong QCD interactions, and 
are expected to be very heavy, say $m_\rho \approx 800 GeV$\cite{eichten86}. 
These heavy technirhos therefore will decouple from the HERA $\empp$ 
collision process.  Although the color-triplet technirhos may be relatively 
light in some new TC models, the stringent Tevatron limits clearly 
ruled out the  vector leptoquark interpretation
for HERA high-$Q^2$ anomaly. We thus will not consider these vector TC  
leptoquarks anymore.
 
As shown in Table 1, the 24 color-triplet PGB's form one isotriplet 
$(P^1_3, P^0_3, P^{-1}_3)$ and isosinglet $P^{'}_3$ and their antiparticles. 
They are the usual
$F=0$ scalar leptoquarks. The $P_3^0$ and $P_3^{'}$ are the mixed states of 
the charge $2/3$ pseudo-Goldstone bosons 
 $P_{U\overline{N}}$ and  $P_{D\overline{E}}$ \cite{ellis81,eichten86}.
At HERA, $P_3^{1}$, $P_3^0$ and $P_3^{'}$ could be produced directly by 
resonant production of  $e^{\pm}p \to LQ$ if they have low masses 
$m_{LQ} < \sqrt{s}$ \cite{berger91}. The  $P_3^{-1}$ is irrelevant to 
HERA experiments because  it could not be produced
by $\empp$ collision process at HERA, and  the Tevatron lower mass 
bounds on $M_S$ also do not apply to $P_3^{-1}$ since   it decays uniquely to
$\overline{\nu}d$ final state. 
 
According to previous studies \cite{eichten80,rs95}, the scalar leptoquarks 
$P_3^{1,0,-1}$ and $P_3^{'}$ receive masses from QCD, 
ETC and electroweak interactions, $m(P_3) \sim 160 G$eV
with a mass splitting of about 10 GeV. But we also know   that 
the TC leptoquark masses could be increased greatly in Walking TC theories 
since the  condensate enhancement \cite{appel87} also enhances the masses of 
TC leptoquarks. We also expect additional 
uncertainties for Multiscale Technicolor Model \cite{lane89} and other
TC models where the TC dynamics is quite different from the QCD.

\begin{table}[htbp]
\begin{center}
\caption{Technicolor leptoquarks in the Farhi-Susskind one generation 
TC model, as given in Table II of ref.[18]  . 
The $SU(3)_c$ index $\alpha$ runs over 1,2,3.}
\label{table1}
\vskip 0.5pc
\begin{tabular}{|c||c||c|c|c|}\hline\hline
States    &Technifermion wave function  &$SU(3)_c$   
&$(I, I_3)$ & Charge   \\  \hline\hline
$P_3^1$, \ \ $\rho_3^1$&  $|U_\alpha\overline{E}>$& 3 & (1,1)& 5/3\\ \hline
$P_3^0$, \ \ $\rho_3^0$&  $(1/ \sqrt{2})
|U_\alpha\overline{N} - D_\alpha \overline{E}>$& 3 & (1,0)& 2/3\\ \hline
$P_3^{-1}$, \ \ $\rho_3^{-1}$&  
$|D_\alpha\overline{N}>$& 3 &(1,-1)& -1/3\\ \hline
$P_3^{'}$, \ \ $\rho_3^{'}$& $(1/ \sqrt{2})
|U_\alpha\overline{N} + D_\alpha \overline{E}>$& 3 & (0,0)& 2/3\\ \hline\hline
\end{tabular}
\end{center}
\end{table}

The gauge interactions of technicolor  leptoquarks with the 
standard model gauge bosons  occur
dynamically through technifermion loops. Their coupling to gauge bosons
$(\gamma, Z^0, W^{\pm}, gluon)$ can be evaluated reliably by using
well-known techniques of current algebra or effective lagrangian
methods. Consequently, the parameter free pair production cross sections
of leptoquarks at $\overline{p}p$ colliders have been calculated at 
leading and next-to-leading order \cite{kramer97}. It is this merit 
that makes it possible for D0 and CDF collaborations to obtain their 
 lower mass bounds on leptoquarks based 
on their negative searches at Tevatron.

The coupling of a TC leptoquark to lepton-quark pairs occurs when the 
technifermion constituents of the leptoquark exchange an ETC gauge boson 
and turn into an ordinary quark and lepton. 
the ''Yukawa'' coupling of leptoquarks is therefore model-dependent and 
currently not known with confidence. 
Following ref.\cite{ellis81},  we rewrite the effective Yukawa
couplings of F=0 scalar TC leptoquarks to lepton-quark pairs before mixing 
as the form of 
\beq
{\cal L}&=&\lambda_{1L} \punb \overline{\nu}_L u_R + 
\pdeb \left[ 
\lambda_{2L} \overline{e}_L d_R -\lambda_{2R}\overline{e}_R d_L \right] 
\nonumber  \\ 
&& + \pueb \left[ \lambda_{3L} \overline{e}_L u_R - \lambda_{3R}
 \overline{e}_R u_L \right]  + 
\lambda_{4L}\pdnb \overline{\nu}_L d_R + h.c.
\label{yc}
\eeq
where the $e, \nu, d,$ and $u$ are vectors $(e, \mu, \tau)$, 
$(\nu_e, \nu_\mu, \nu_\tau)$, $(d, s, b)$ and $(u, c, t)$ respectively,
and the effective Yukawa couplings $\lambda_i$ in eq.[\ref{yc}] 
generally depend on the 
Technicolor and Extended Technicolor dynamics and therefore could 
vary over a large range. 
If we assume, according the general sense, 
that the Yukawa coupling is something  
like $\lambda    \approx (m_q +m_l)/F_\pi$ ( the $F_\pi$ is the 
Goldstone boson decay constant and $F_\pi =123 GeV$ in the QCD-like OGTM), 
the couplings will be
very small for light $(u,d)-e^\pm $ pairs. Although the low energy 
constraints could  be avoided automatically by such kinds of 
couplings \cite{roger87}, but numerical  estimation shows that 
the relevant couplings ( i.e. $\lambda_{e^+u}$, $\lambda_{e^+d}$ and 
$\lambda_{\overline{\nu}u}$ ) are too small to provide an adequate contribution
to the $\epp$ collision process at HERA. 
For $ \lambda_{e^+u}=\lambda_{e^+d}= \lambda_{\overline{\nu}u}
\approx m_u/F_{\pi} \approx 10^{-4}$, for example, 
the neutral current cross-section from $\ptp$ 
is smaller than $1\; fb$ assuming $m(P_3^{'})=200\; GeV$ and $\beta =0.7$.

On the other hand, in the popular BRW leptoquark scenario
\cite{buchmuller87}, 
the interactions  of leptoquarks with  
quark-lepton pairs were  described by an ``effective'' Lagrangian
\cite{buchmuller87} involving scalar and vector leptoquarks with 
general $SU(3)_C\times SU(2)_L\times U(1)_Y$
invariant Yukawa couplings $\lambda_{L,R}^{i,j}$, constrained  by low-energy
experiments.  The $F=0$ scalar TC leptoquarks in Table 1 are indeed the 
same kinds of leptoquarks as the {\cal R} and {\cal \~R}  leptoquarks in 
ref.\cite{buchmuller87} and couple to quark-lepton pairs in the similar way. It 
is  therefore reasonable for us to assume that the Yukawa couplings of 
TC leptoquarks are also the 
dimensionless parameters to be determined by experiments.  They are 
constrained by low energy experiments in the same way as that for ordinary 
leptoquarks: 

\begin{quotation}

(a). baryon- and lepton number conserving; 

(b). no generation mixing, a given leptoquark couple only to one generation
of fermions and only via $\lambda_L$ or only via $\lambda_R$;

(c). $(\lambda/\sqrt{4\pi \alpha}) < 0.17(M_{LQ}/200GeV)^2$
\cite{leurer}. 
 
\end{quotation}

As for the explicit mixing
patterns of $\ptp$ and $\pto$, we assume that
\beq
\begin{array}{ccl}
\left( \begin{array}{c} \ptp \\ \pto \end{array} \right)
&=&
\left( \begin{array}{c} \cos \theta \ \ \ \sin \theta \\
-\sin \theta \ \ \ \cos \theta \end{array} \right)
\left( \begin{array}{c} \punb \\ \pdeb \end{array} \right)
\end{array}
\eeq

After mixing, the effective Yukawa couplings of the  $\ptp$ and $\pto$
to lepton-quark pairs are
\beq
{\cal L} &=& \ptp \left[
\lambda_{2L} \sin \theta \overline{e}_L d_R -\lambda_{2R}\sin \theta
 \overline{e}_R d_L +
\lambda_{1L} \cos \theta \overline{\nu}_L u_R \right] \nonumber \\
&+& \pto \left[ \lambda_{2L} \cos \theta \overline{e}_L d_R -\lambda_{2R}
\cos \theta \overline{e}_R d_L -
\lambda_{1L} \sin \theta \overline{\nu}_L u_R \right] + h.c.
\label{yceff}
\eeq
where the $\lambda_{1L,R}$ and $\lambda_{2L,R}$ are the effective Yukawa
couplings before mixing as given in eq.[\ref{yc}]. For $\ptp$ and $\pto$,
the branching ratios to the $e^+d$ final state are
\beq
BR(\ptp \to e^+d)&=&\frac{(\lambda_{2L}^2 + \lambda_{2R}^2)
 \sin^2 \theta}{(\lambda_{2L}^2 + \lambda_{2R}^2) \sin^2 \theta
+ \lambda_{1L}^2 \cos^2 \theta}, \\
BR(\pto \to e^+d)&=&\frac{(\lambda_{2L}^2 + \lambda_{2R}^2)
 \cos^2 \theta}{ (\lambda_{2L}^2 + \lambda_{2R}^2) \cos^2 \theta
+ \lambda_{1L}^2 \sin^2 \theta}.
\eeq
The size  of the branching ratios depend on  the values of $\lambda_1$,
$\lambda_2$ and $\theta$. For the sake of
simplicity, we assume that $\theta = \pi/4$ and $\lambda_R = 0$
\footnote{If $\lambda_R\neq 0$ and $\lambda_L = 0$ the $\ptp$ and
$\pto$ could not contribute to CC process as being shown in section 3, 
we therefore assume implicitly that
$\lambda_L \neq 0$ and  $\lambda_R=0$ in the following calculations.}.

In Table 2 we list the  Yukawa couplings of all F=0 scalar TC leptoquarks to 
first generation lepton-quark  pairs, 
and the couplings of  {\cal R} and {\cal $\tilde{R}$} \cite{buchmuller87} 
to lepton-quark pairs are also included as a comparison. 

\begin{table}[htbp]
\begin{center}
\caption{The Yukawa couplings of the F=0 scalar Technicolor leptoquarks 
to first generation quark-lepton pairs. The subscripts $L, R$ of the coupling 
refer to the lepton chirality. The couplings  {\cal R} and
 {\cal $\tilde{R}$}  to lepton-quark pairs  are listed as a comparison. }
\label{table2}
\vskip 0.5pc
\begin{tabular}{|c||c||c|c|c||c|c|}\hline\hline
Channel    & $P_3^{'}$ & $P_3^1$ & $P_3^0$ & $P_3^{-1}$ & 
${\cal R}$&  {\cal $\tilde{R}$ }  \\  \hline\hline
$e^-_{L,R} \bar d$& $\lambda_{2L}$,   $\; -\lambda_{2R}$   
& -  & $\lambda_{2L}$,  $\; -\lambda_{2R}$& -& $-h_{2R}$&
$\tilde{h}_{2L}$ \\ \hline
$e^-_{L,R} \bar u$&  -  & $ \lambda_{3L}$,   $\; -\lambda_{3R}$  
& -   &- & $h_{2L,R}$&- \\ \hline
$\nu_L \bar u$   & $\lambda_{1L}$  &- & $-\lambda_{1L}$&- 
& $h_{2L}$&- \\ \hline
$\nu_L \bar d$   &-  & -  & - & $\lambda_{4L}$& - 
& $\tilde{h}_{2L}$ \\ 
 \hline\hline
\end{tabular}
\end{center}
\end{table}

The TC leptoquarks
 $\ptp$,   $\pto$ and $P_3^1$ would contribute to NC and/or CC processes
in the similar  way if they were not heavy, but they could not be
light simultaneously because of the stringent Tevatron mass bounds 
\cite{cdf210,d0225,kramer972}. In order to compare with the HERA 
data quantitatively, we consider the following three typical cases in detail:

\begin{itemize}

\item[Case-1].   
Assuming  $m(P_3^1)=225 GeV$, $ m(P_3^{'}) = m(P_3^0)=280 \sim 300 GeV$,
i.e., only the charge 5/3 TC leptoquark $P_3^1$ contribute effectively; 
If we apply the combined Tevatron limit, $m(P_3^1)=240 GeV$ is still allowed 
\cite{kramer972};

\item[Case-2].  
Assuming  $m(P_3^{'})=200 GeV$, $ m(P_3^1) = m(P_3^0)=
280 \sim 300 GeV$, 
i.e., only the lighter mixed state  $P_3^{'}$ contribute effectively; For
this case, $\beta_{NC} \leq 0.7 (0.5)$ is allowed by CDF and D0 (combined)
limit \cite{cdf210,d0225,kramer972};

\item[Case-3].  
Both $\ptp$ and $\pto$ contribute effectively. If we assume that 
$m(P_3^{'})=200 GeV$, $m(P_3^0)=225 GeV$ and $m(P_3^1)=280 \sim 300 GeV$, 
the value of branching ratios $\beta_1=\breda$ and $\beta_2=\bredb$ 
will be constrained strongly by the Tevatron upper limits on the 
pair production cross section of scalar leptoquarks. For $\beta_1=0.6$, 
$\beta_2 \leq 0.5$ is allowed by D0 $95\% C.L.$ upper limit of 
$\sigma \leq 0.078 pb$. For $\beta_1=0.4$, $\beta_2 \leq 0.4$ is 
allowed by combined Tevatron upper limit of $\sigma \leq 0.04 pb$. 

\end{itemize}

\section*{3. Contributions from TC leptoquarks}

In this  section we will calculate  the contributions to the
$\empp$ collision process from those F=0 scalar TC leptoquarks. 
 In numerical calculations,
we use the LEPTO 6.5 \cite{lepto65} program and the  MRS96 parton
distribution functions \cite{mrs96}
with the inclusion of high order QED and QCD corrections. For all relevant
masses, decay widths or coupling constants such as $M_Z$, $M_W$, $\alpha$,
$G_F$, etc., we use the default values as given in PYTHIA 5.724/JETSET
7.410 \cite{pythia}. 

The TC leptoquarks $(P_3^1, P_3^0, P_3^{'})$ can be produced directly in
$\epp$ deep inelastic scattering from a u or d valence quark
in proton  as illustrated in Figs.(1a,1c,1e), if 
their masses  are  smaller than the $\empp$ center of
mass energy $\sqrt{s}$ \cite{cashmore85}.
The leptoquarks  will also contribute indirectly by u-channel 
exchange as shown in
Figs.(1b, 1d, 1f). The TC leptoquarks can also be produced
through  ``gluon fusion'' in which the incoming positron annihilates
with the quark from the virtual $q\overline{q}$ pair of a gluon. We
will not consider the case of ``gluon fusion'' since the corresponding
production rate is unobservably small at HERA \cite{cashmore85,djouadi90}.
we at first present the relevant formulae being 
used in the numerical  calculations.

\subsection*{3.1 Electroweak and new physics cross sections}

At HERA $\empp$ collider, the reaction $e^{\pm} + p \to l^{\pm} + X$ 
is expected to occur through the subprocess
\beq
e^{\pm} + q_1 \to l^{\pm} + q_2
\eeq
where the $q_1$  is an initial state quark in the proton, $q_2$ is the final 
state quark. In the framework of the SM, $l^{\pm}=e^{\pm}$, $\nu$ or 
$\overline{\nu}$ for NC and CC 
processes respectively. In new physics models beyond the SM, the $l^{\pm}$ 
could  be all three generation leptons \cite{syang96}. But we here 
consider only the first generation fermions.

In leading order, 
the differential NC cross-section for an incoming polarized 
electron  beam colliding with unpolarized proton beam is given by
\cite{lepto65}
\beq
\frac{d^2\; \sigma_{NC}(e^-_{L,R} p)}{dx\;dQ^2}= 
\frac{2\pi \alpha^2}{xQ^4}
\left[ (1+ (1-y)^2) F_2^{L,R}(x, Q^2) 
+(1-(1-y)^2) x F_3^{L,R}(x, Q^2)\right]
\label{nccs}
\eeq
where the structure functions $F_{2,3}^{L,R}$ are the form of 
\beq
F^{L,R}_2(x,Q^2)&=& \sum_q x\left[ q(x,Q^2) +
\overline{q}(x,Q^2) \right] A^{L,R}_q,  \\
x F^{L,R}_3(x,Q^2)&=& \sum_q x\left[ q(x,Q^2) - 
\overline{q}(x,Q^2) \right] B^{L,R}_q,  
\eeq
with the coefficient functions  
\beq
A^{L,R}_q(Q^2)&=& e_q^2 - 2e_q(v_e \pm a_e)v_q P_Z 
+ (v_e \pm a_e)^2 (v_q^2 + a_q^2 ) P_Z^2, \\
B^{L,R}_q(Q^2) &=& \mp 2e_q (v_e \pm a_e)a_q P_Z 
\pm 2 (v_e \pm a_e)^2 v_qa_q P_Z^2
\eeq
where $e_q$ is the electric charge of quarks( $e_e=-1$), $v_f=(I_{3f} 
- 2e_f \sin^2\theta_W)/\sin 2\theta_W$ and $a_f=I_{3f}/\sin 2\theta_W$ are the 
vector and axial vector electroweak couplings, and 
$P_Z=Q^2/(Q^2 + M_Z^2)$. For incoming $e^+_{L,R}$ beams, the corresponding 
cross sections are obtained from the above formulae by the replacements
\beq
F_2^{L,R} \longrightarrow F_2^{R,L}, \ \ 
x F_3^{L,R} \longrightarrow - x F_3^{R,L}.
\eeq

For the charged current $e^-p$ ($e^+p$) collision process, only the polarized  
$e_L^-$ ($e_R^+$) beam contribute.  If we  consider only 
four massless quark flavours $(u, d, s, c)$ and use the unitarity relation of 
CKM matrix, the differential cross section for incoming $e_L^-$ and $e_R^+$  
beams   colliding with unpolarized proton are  given respectively by
\cite{lepto65}
\beq
\frac{d^2\sigma_{CC}(e_L^- p)}{dx\;dQ^2} \approx
\frac{G_F^2}{\pi} ( 1+ \frac{Q^2}{M_W^2})^{-2}
\left[ (u + c) + (1-y)^2 (\overline{d} + \overline{s}) \right]\\
\frac{d^2\sigma_{CC}(e_R^+ p)}{dx\;dQ^2} \approx
\frac{G_F^2}{\pi} ( 1+ \frac{Q^2}{M_W^2})^{-2}
\left[ (\bar u + \bar c) + (1-y)^2 (d + s) \right]
\eeq

The explicit formulae for the leptoquark contributions to both NC and CC 
$\empp$ collision processes are rather simple 
If we neglect all  interference terms such as that  
between the SM amplitudes and leptoquark amplitudes. These interference 
terms are in general very small in the high-$Q^2$ region. 
In the mass region ($m_{LQ} < \sqrt{s}=300
GeV$) considered here, the s-channel  will dominate.

For a F=0 scalar TC leptoquark, the neutral current  
leptoquark differential cross sections
for an incoming polarized $e^{\pm}$ beam colliding with an unpolarized 
proton are
\beq
\frac{d^2 \sigma_{NC}(e^-_{L}p)}{dx\;dQ^2}
=   \frac{2\pi \alpha^2}{xQ^4}  \frac{F^i_L( F^f_L + F^f_R)}{4} 
\left[   \frac{\hat{t}^2 x \overline{q}(x, Q^2)}{ [ \hat{s} - m_{LQ}^2 ] ^2  
+ m_{LQ}^2 \Gamma^2_{LQ} } 
+  \frac{\hat{u}^2 y^2 x q(x,Q^2)}{[\hat{u}-m_{LQ}^2 ]^2 }  
\right] 
\label{elmnc} \\
\frac{d^2 \sigma_{NC}(e^+_{R}p)}{dx\;dQ^2} =
\frac{2\pi \alpha^2}{xQ^4} \frac{F^i_L( F^f_L + F^f_R)}{4}  \left[
\frac{\hat{t}^2 x q(x, Q^2)}{ [ \hat{s} - m_{LQ}^2 ]^2
+ m_{LQ}^2 \Gamma^2_{LQ} } +  
\frac{\hat{u}^2 y^2 x \overline{q}(x,Q^2)}{[\hat{u}-m_{LQ}^2 ]^2 } \right] 
\label{erpnc}
\eeq
where the $\hat{s}, \hat{t}$ and $\hat{u}$ are the Mandelstam variables,
defined as $\hat{s} = sx$, $\hat{t}= - Q^2 = -sxy$ and $\hat{u} = -\hat{s}
 + Q^2= -xy(1-y)$. And the $F_{L,R}$ are the redefined couplings and the 
superscripts $i$ and
$f$ refer to the couplings at the production and the decaying vertices,   
\beq
F_{L,R}= \lambda^2_{L,R}/( 4\pi \alpha) 
\label{flr} 
\eeq
with electroweak coupling $F_{EW}=1$. We 
neglected the interference terms between the SM amplitudes and leptoquark 
amplitudes, since such interference  terms are  very small, 
say less than 1 $fb$ , for $Q^2 >
10000 GeV^2$. The cross sections for incoming $e^-_R$ and $e_L^+$ beams 
are obtained by exchanging
the $L$ and $R$ in eq.[\ref{elmnc}] and eq.[\ref{erpnc}], respectively. 
The low energy constraints on effective Yukawa couplings can be rewritten 
as 
\beq
F_L \leq 0.03 \left( \frac{m_{LQ}}{200 GeV}\right)^4
\label{fl}
\eeq
and we had assumed that $F_R \equiv 0$.

For charged current  process $\empp \rightarrow \overline{\nu} (\nu) X$, 
the F=0 scalar leptoquark interactions do not interfere with the standard
model DIS, and 
the leptoquark  differential cross section
for an incoming polarized $e^{\pm}$ beam colliding with an unpolarized
proton are \cite{st96}
\beq
\frac{d^2 \sigma_{CC}(e^-_{L}p)}{dx\;dQ^2}&=&
\frac{2\pi \alpha^2}{xQ^4}  \frac{F^i_LF^f_L}{4}  \left[
\frac{\hat{t}^2 x \overline{d}(x, Q^2)}{ [ \hat{s} - m_{LQ}^2 ] ^2
+ m_{LQ}^2 \Gamma^2_{LQ} } +  \frac{\hat{u}^2 y^2 x u(x,Q^2)}{[\hat{u}-
m_{LQ}^2 ]^2 }  \right] \\
\frac{d^2 \sigma_{CC}(e^-_{R}p)}{dx\;dQ^2}&=&
\frac{2\pi \alpha^2}{xQ^4}  \frac{F^i_R F^f_L}{4}  \left[ 
\frac{\hat{t}^2 x \overline{d}(x, Q^2)}{ [ \hat{s} - m_{LQ}^2 ] ^2
+ m_{LQ}^2 \Gamma^2_{LQ} } +  \frac{\hat{u}^2 y^2 x u(x,Q^2)}{[\hat{u}-
m_{LQ}^2 ]^2 }  \right] \\ 
\frac{d^2 \sigma_{CC}(e^+_{L}p)}{dx\;dQ^2}&=&
\frac{2\pi \alpha^2}{xQ^4}  \frac{F^i_L F^f_R}{4}  \left[
\frac{\hat{t}^2 x d(x, Q^2)}{ [ \hat{s} - m_{LQ}^2 ] ^2
+ m_{LQ}^2 \Gamma^2_{LQ} } +  
\frac{\hat{u}^2 y^2 x \overline{u}(x,Q^2)}{[\hat{u}-
m_{LQ}^2 ]^2 }  \right] \\ 
\frac{d^2 \sigma_{CC}(e^+_{R}p)}{dx\;dQ^2}&=&
\frac{2\pi \alpha^2}{xQ^4}  \frac{F^i_L F^f_L}{4}  \left[
\frac{\hat{t}^2 x d(x, Q^2)}{ [ \hat{s} - m_{LQ}^2 ] ^2+ 
m_{LQ}^2 \Gamma^2_{LQ} } +
\frac{\hat{u}^2 y^2 x \overline{u}(x,Q^2)}{[\hat{u}-m_{LQ}^2 ]^2 }  \right] 
\eeq
It may be seen that there could  be no TC leptoquark contribution to CC 
processes if $F_L=0$, while only  the incoming $e_L^-$ ($e^+_R$) beam 
contributes to the CC $\emp$ ($\epp$) collision  
process if $F_L \ne 0$ and $F_R =0$.

\subsection*{3.2 Case-1, NC contribution from $P_3^1$  }

As shown in Fig.1a, the charge $5/3$ TC leptoquark $P_3^1$ can be produced by 
$e^+ u$ fusion and will decay uniquely back  to $e^+ u$ lepton-quark pair. 
It can not contribute to the CC cross section, but 
could  provide a rather large contribution to NC process if it was light. 

As discussed in Section 2, $m(P_3^1) =225$ (240) GeV is allowed by D0 
(combined Tevatron) 
limit when other TC leptoquarks are heavy and effectively decouple.

Table 3 shows the integrated NC cross 
sections from the $P_3^1$, assuming 
$F_{3L}=\lambda_{3L}^2/(4\pi \alpha)
=(0.005, 0.01, 0.02), 
F_R=0$ and $m(P_3^1)=200-300$ GeV, 
respectively. For $m(P_3^1)=225 GeV$ and a left-handed chiral coupling 
of $0.008$,  the $P_3^1$ itself 
can  provide the required  extra NC cross section, 
say $\Delta \sigma_{NC} \approx 0.2 pb$ for $Q^2_{min}=15000 GeV^2$,  
to explain the HERA NC anomaly. This result is consistent with that 
of previous similar studies \cite{lq1}. 
For $m(P_3^1)=240 GeV$, the required extra contributions can still be 
achieved for $F_{3L}=0.02$.
 For more heavier 
$P_3^1$, its contribution decreases rapidly and can be neglected for 
$m(P_3^1) \ge 280 GeV$.  
But the key problem for $P_3^1$ is that it could  not contribute to CC process. 
A relatively light $P_3^1$ therefore will  be excluded 
 if the excess of CC events with high-$Q^2$ is finally confirmed by 
HERA data. 

\begin{table}[htbp]
\begin{center}
\caption{The neutral current  cross sections from TC leptoquark $P_3^1$ with 
$Q^2_{min}=15000, 25000 GeV^2$, 
assuming $F_{3L}=(0.005, 0.01, 0.02), F_R=0$ and $m(P_3^1)=200-300$ GeV,
respectively. All cross sections in $pb$}
\label{table3}
\vskip 0.5pc
\begin{tabular}{|c||c|c|c||c|c|c|}\hline
Mass &\multicolumn{3}{c||}{$Q^2_{min}=15000 GeV^2$}&
      \multicolumn{3}{c|}{$Q^2_{min}=25000 GeV^2$}\\ \hline
(GEV)&$F_{3L}=0.005$&$0.01$&$0.02$&$F_{3L}=0.005$&$0.01$
&$0.02$\\ \hline
200&0.324&0.651&1.297&0.192&0.384&0.767 \\
220&0.145&0.293&0.582&0.100&0.201&0.401\\
240&0.047&0.096&0.189&0.035&0.071&0.142\\
260&0.009&0.019&0.035&0.007&0.014&0.028\\
280&0.0005&0.002&0.004&0.0003&0.001&0.002\\
300&$10^{-6}$&0.0007&$10^{-5}$&0&0.0002&0.0002\\ \hline
\end{tabular}
\end{center}
\end{table}

\subsection*{3.3 Case-2, the NC and CC contributions from single $\ptp$}

For the assumed mass spectrum of Case-2, only  $\ptp$, the lighter mixed 
state, contribute to both NC and CC $\epp$ collision processes effectively. 
For $m(\ptp)=200 GeV$, $\beta_{NC} \leq 0.7 $ (0.5) is allowed by D0 
(combined) limit. As for the Yukawa coupling $F_{1L}$, we have 
\beq
F_{1L}= F_{2L} (1-\beta_{NC})/\beta_{NC}
\eeq
Assuming $F_{2L}=\lambda_{2L}^2/(4\pi \alpha)=0.02$ and 
$\beta_{NC}=(0.7,0.6,0.5)$, we have $F_{1L}=0.009, 0.013$ 
and $0.02$ respectively. 

Table 4 shows the 
contributions to NC cross sections from $\ptp$, assuming $F_{2L}=0.02$
and $\beta_{NC} = 0.7, 0.6 $ and $0.5$ respectively. As a comparison, 
we also list the preliminary 1994-97 combined ZEUS and H1 NC cross 
sections with different $Q^2_{min}$ cuts\cite{straub}, 
and the corresponding standard model predictions for NC cross sections.

\begin{table}[htbp]
\begin{center}
\caption{The integrated NC cross sections from TC leptoquark $P_3^{'}$,
assuming $m(\ptp)=200 GeV$, $F_{2L}=0.02$. The $\slq^{a}$, $\slq^{b}$ and 
$\slq^{c}$ corresponding to $\beta_{NC} =0.7, 0.6$ and $0.5$ respectively. 
The second and third columns show the combined H1 and ZEUS   NC 
data  with $Q^2_{min}$ cuts. All cross sections in $pb$.}
\label{table4}
\vskip 0.5pc
\begin{tabular}{|c||c|c|c||c|c|c|}\hline
$Q^2_{min}$ &\multicolumn{2}{c|}{ H1 + ZEUS }& & 
 \multicolumn{3}{c|}{ $\ptp$ Contributions} \\ \hline
($GeV^2$) &$N_{obs}$& $\sigma_{obs}$&$\sigma_{sm}$
&$\slq^{a}$&$\slq^{b}$& $\slq^{c}$ \\ \hline
2500 &724 $^{(a)}$ &$43.3^{+4.6}_{-3.9}$&45.7&0.337  &0.289   &0.241 \\
5000&193 + 326  $^{(b)}$& $10.7 \pm 0.7$& 10.6&0.311&0.267&0.222 \\ 
10000&31 + 50 $^{(b)}$& $1.70 ^{+ 0.23}_{-0.20}$&1.79&0.263&0.226&0.188\\
15000&18+ 18 $^{(b)}$&$0.71 ^{+0.14}_{-0.12}$&0.49&0.217&0.186&0.155  \\
20000&7 + 7 $^{(b)}$&$0.30^{+0.092}_{-0.076}$&0.161&0.172&0.147&0.123  \\
25000&4 + 3 $^{(b)}$&$0.16^{+0.069}_{-0.053}$&0.059&0.128&0.110&0.091  \\
30000&2+ 2 $^{(b)}$&$0.098^{+0.059}_{-0.042}$&0.023&0.085&0.073&0.060  \\
35000&2$^{(c)}$&$0.060^{+0.059}_{-0.037}$&0.0091&0.042&0.036&0.030  \\
40000&1$^{(c)}$&$0.032^{+0.044}_{-0.023}$&0.0036&0.00004 $^{(d)}$ 
&0.00004 $^{(d)}$&
0.00004 $^{(d)}$ \\ \hline
\end{tabular}
\end{center}
\noindent
(a). ZEUS 1994-97 $e^+p$ NC data, ${\cal L} = 33.5 pb^{-1}$;\\
(b). Combined H1 and ZEUS 
1994-97 $e^+p$ NC data, ${\cal L } = 57.2 pb^{-1}$; \\
(c). H1 1994-97 $e^+p$ NC data, ${\cal L } = 23.7 pb^{-1}$;\\
(d). For $m(\ptp)=215 GeV$ and $\beta_{NC} = 0.7, 0.6$ and $0.5$, 
the $\slq$ is 0.020, 0.017 and 0.014 respectively.
\end{table}

As shown in  Table 5, the F=0 scalar TC leptoquark $\ptp$ with $m(\ptp)=200 
GeV$ can indeed provide the 
required extra contributions to the CC process. 
For $Q^2_{min}=10000 GeV^2$, for example, 
$\scc^{a,b,c}= (0.381, 0.509, 0.636)\; pb$, assuming
$F_{2L}=0.02$, $F_R=0$ and $\beta_{CC} = 0.3, 0.4$ and
$ 0.5$ respectively. The SM predictions and the
preliminary CC results from 1994-97 H1 and ZEUS $e^+p$ data \cite{straub}
are also included in Table 5 as a comparison.

At HERA, the initial and final state particles in leptoquark-mediated 
interactions are identical to those in NC and CC DIS. Events due to 
leptoquark-mediated interactions are therefore expected to be 
identical to those from DIS processes. 
Like the DIS CC events, The TC leptoquark CC events will be 
experimentally very clean and can be selected with similar high  efficiency. 
In order to estimate roughly  how much extra CC
contributions are required from the new physics sources, based on the
 current H1 and ZEUs data, we define $\triangle \sigma_{CC}$
the required extra CC cross section, as
\beq
\triangle \sigma_{CC}= \frac{N_{obs}-N_{exp}}{N_{exp}}\sigma_{SM} \pm
\frac{\delta N_{exp}}{N_{exp}}\sigma_{SM}
\label{dcc}
\eeq
where the $N_{obs}$ and $N_{exp}$ are the numbers of observed and expected
CC events  by H1 and/or ZEUS collaboration respectively, and the
$\sigma_{SM}$ is the SM CC cross section.

\begin{table}[htbp]
\begin{center}
\caption{The integrated CC cross sections from TC leptoquark $P_3^{'}$ with
different $Q^2_{min}$ cuts,
assuming $m(\ptp)=200 GeV$, $F_{2L}=0.02$. The $\scc^{a}$, $\scc^{b}$ and
$\scc^{c}$ corresponding to $\beta_{CC} =0.3, 0.4$ and $0.5$ respectively.
The ZEUS and H1 preliminary 1994-97 $e^+p$ CC data included. 
All cross sections in $pb$.}
\label{table5}
\vskip 0.5pc
\begin{tabular}{|c||c|c|c||c|c|c|}\hline
$Q^2_{min}$ &\multicolumn{2}{c|}{ H1 + ZEUS }& &
 \multicolumn{3}{c|}{ $\ptp$ Contributions} \\ \hline
($GeV^2$) &$N_{obs}$& $N_{expect}$&$\sigma_{sm}$
&$\scc^{a}$&$\scc^{b}$& $\scc^{c}$ \\ \hline
1000  &455 $^{(a)}$& $419 \pm 36$&16.47&0.391&0.521&0.651 \\
2500 &61 $^{(b)}$ & $56.3 \pm 9.4$&7.32&0.389&0.519&0.649  \\
5000  &43 $^{(b)}$ &$34.7 \pm 6.9$&2.54&0.381&0.509&0.636  \\ 
10000&13 + 15  $^{(c)}$& $17.7 \pm 4.3$ &0.493&0.351&0.468&0.584  \\
15000&6 + 5 $^{(c)}$&$4.9 \pm 1.7$&0.127&0.306&0.408&0.510  \\
20000&4 + 1$^{(c)}$&$1.7 \pm 0.7$&0.037&0.253&0.338&0.442  \\
30000 &1 $^{(a)}$&$0.034 ^{+0.038}_{-0.018}$&0.004&0.133&0.177&0.221  \\ \hline
\end{tabular}
\end{center}
\noindent
(a). ZEUS 1994-97 $e^+p$ CC data, ${\cal L} = 33.5 pb^{-1}$;\\
(b). H1 1994-97 $e^+p$ CC data, ${\cal L } = 23.7 pb^{-1}$;\\
(c). Combined H1 and ZEUS 1994-97 $e^+p$  CC data, ${\cal L } = 57.2 pb^{-1}$.
\end{table}

Table 6 shows the $\triangle \sigma_{CC}$ from the preliminary 1994-97 
H1 and ZEUS $e^+p$ data. 
As a simple  estimation, we neglected the difference between the $Q^2_h$ and 
$Q^2_{JB}$ used by H1 and ZEUS respectively. 
The data show clearly a
need for positive extra contribution $\Delta \sigma_{CC}$, especially 
for CC events with very high-$Q^2$. For $Q^2_{min} = 10000$, $25000 GeV^2$, the 
deviation is $+2.4\sigma$ and $+4.7\sigma$ respectively. For $Q^2_{min} 
= 30000 GeV^2$, ZEUS observed 1 event while the SM expectation is only 
$0.034 ^{+0.038}_{-0.018}$. Inclusion of the TC leptoquark contribution 
can achieve a very good agreement  between the data and the theoretical 
expectations.

For $\emp$ collision process, the F=0 scalar TC leptoquarks 
$(\overline{P}_3^1, \overline{P}_3^0, \overline{P}_3^{'})$ could be produced 
directly by the s-channel $e^-\bar u$ and/or $e^- \bar d$ ``fusion". But
the relevant cross sections are strongly suppressed by the smallness of 
the parton density of sea quarks in a proton. Assuming $m(\ptp)=200 GeV$, 
$F_{2L}=0.02$, $F_{2R}=0$, $\beta_{NC}=0.7$ and $\beta_{CC}=0.3$, 
the NC cross section due to $\ptp$ is only about $0.01 pb$ (less 
than $1\%$ of the corresponding SM contribution) for $Q^2_{min}=15000 GeV^2$, 
while the CC cross section is only $0.02 pb$ (less than $0.6\%$ of the 
corresponding SM cross section) for $Q^2_{min}=10000 GeV^2$.
The contributions to $\emp$ collision process from F=0 scalar TC leptoquarks 
are indeed very small and can be neglected safely.

\begin{table}[htbp]
\begin{center}
\caption{The required extra CC cross section $\triangle \sigma_{CC}$, and
the $\ptp$ contributions with different $Q^2_{min}$ cuts,
assuming $m(\ptp)=200 GeV$, $F_{2L}=0.02$. The $\scc^{a,b,c}$  
 corresponding to $\brnua =0.3$, $0.4$ and $0.5$ respectively.
The ZEUS and H1 preliminary 1994-97 $e^+p$ CC data included.
All cross sections in $pb$.}
\label{table6}
\vskip 0.5pc
\begin{tabular}{|c|c|c|c|c|c|c|c|c|}\hline
$Q^2_{min}$ &\multicolumn{2}{c|}{ H1 + ZEUS }& & & &
 \multicolumn{3}{c|}{ $\ptp$ Contributions} \\ \hline
($GeV^2$) &$N_{obs}$& $N_{expect}$&$\delta$&$\sigma_{sm}$
&$\triangle \sigma_{CC}$&$\scc^{a}$& $\scc^{b}$& $\scc^c$ \\ \hline
1000  &455&$419 \pm 36$     &$+1\sigma$ &16.47  & $1.4 \pm 1.4$
&0.391&0.521&0.651 \\ 
2500  &61 & $56.3 \pm 9.4$  &$+0.5\sigma$ &7.32 &$0.61\pm 1.22$ 
&0.389&0.519&0.649  \\
5000  &43 &$34.7 \pm 6.9$   &$+1.2\sigma$&2.54  &$0.61\pm 0.51$
 &0.381&0.509&0.636  \\
10000 &28 & $17.7 \pm 4.3$  &$+2.4\sigma$&0.493 &$ 0.287\pm 0.12$ 
&0.351&0.468&0.584  \\
15000 &11 &$4.9 \pm 1.7$    &$+3.6\sigma$&0.127 &$0.158\pm 0.04$
&0.306&0.408&0.510  \\
20000 &5  & $1.7 \pm 0.7$   &$+4.7\sigma$&0.037 &$ 0.072\pm 0.015$
&0.253&0.338&0.442  \\
30000 &1  &$0.034 ^{+0.038}_{-0.018}$&$+25\sigma$ 
&0.004&$ 0.114^{+0.004}_{-0.002}$&0.133&0.177&0.221  \\ \hline
\end{tabular}
\end{center}
\end{table}

From Tables (4, 5, 6)   three observations are in order.
First, even one single scalar TC leptoquark $\ptp$
can account for both excesses of NC and CC events with very high $Q^2$
at HERA simultaneously, while all relevant parameters are still within the 
 region allowed by experiments. For the neutral current process,  the 
total cross section, $\sigma_{SM}$ plus $\slq$, reproduce the observed NC 
cross sections  
within the $1-\sigma$ experimental error for the whole range of $Q^2= 2500 - 
40000 GeV^2$. The same is true for the charged current process. For 
$\brnua \ge 0.3$ a single  $\ptp$ with $m(\ptp)=200-220 GeV$ can 
provide an adequate extra cross section to explain the excess of CC events with 
very high-$Q^2$. Secondly, the relative strength of NC and CC contributions 
from $\ptp$ depends on the ratio of the corresponding branching ratio to 
$\epd$ and $\nubu$. The ``best'' choice is apparently $\breda \approx 0.7$. For 
$\breda =0.5$, the NC contribution seems to be a little inadequate while the 
CC contribution seems a bit larger than that required. Finally, the effective
Yukawa coupling of TC leptoquark to quark-lepton pairs must be left-handed, 
i.e., $F_L \neq 0$ and $F_R=0$. Otherwise, there will be no any charged 
current contribution from  the F=0 scalar TC leptoquarks.

\subsection*{3.4 Case-3, contributions from both $\ptp$ and $\pto$}

As discussed in ref.\cite{bassler},
the HERA anomaly is complicated by the
fact that the invariant mass distributions of the event samples of the H1
and ZEUS collaborations are quite different. For the 1994-96 $e^+p$ data,
the 7 H1 NC events clustered at $M \approx 200 GeV$ with the average mass of
$M_e^{avg}=200 \pm 2.6 GeV$ and $M_\omega^{avg}=199 \pm 2.5 GeV$,
while the 5 ZEUS NC events clustered at $M \approx 220 GeV$ 
with the average mass
of $M_{DA}^{avg}=226 \pm 9 GeV$ and $M_\omega^{avg}=216 \pm 7 GeV$.
Bassler and Bernardi claimed that the H1 and ZEUS samples are
concentrated at significantly different mass values and this splitting
cannot be accounted for either by ISR or by detector effects.  B.Straub also
reaches a similar conclusion \cite{straub}: It is unlikely that both excesses
observed  by H1 and ZEUS could be caused by a single narrow resonance!
A natural
and reasonable solution  is that: there are two F=0 scalar
TC leptoquarks with a moderate mass splitting of about 25 GeV, and they both  
contribute to $\epp$ collision processes effectively. That is
the motivation for us to consider the Case-3 in detail.

As given in Table 2, the TC leptoquark $\ptp$ and $\pto$ have the same 
coupling $\lambda_{2L}$ to $e^+d$ pair, while  their couplings to $\bar \nu d$ 
are same in size but with opposite sign. Consequently, the interference terms 
will be constructive to NC cross section but destructive to CC cross section.
For all possible interference terms, only the one between two s-channel 
amplitudes, $2 {\cal R}e(M^S_{1}M^{S*}_{2})$,  
may be important. The differential 
cross sections from this interference term are
\beq
\frac{d^2 \sigma_{NC}(e^-_{L}p)}{dx\;dQ^2}
=   \frac{2\pi \alpha^2}{xQ^4}  \frac{F_{2L}F_{1L}}{4}
\left[ 2 \hat{t}^2 D_{12} x \overline{d}(x, Q^2) \right]
\label{int1} \\
\frac{d^2 \sigma_{NC}(e^+_{R}p)}{dx\;dQ^2}
=   \frac{2\pi \alpha^2}{xQ^4}  \frac{F_{2L}F_{1L}}{4}
\left[ 2 \hat{t}^2 D_{12} x d(x, Q^2) \right]
\label{int2} \\
\frac{d^2 \sigma_{CC}(e^-_{L}p)}{dx\;dQ^2}
=   \frac{2\pi \alpha^2}{xQ^4}  \frac{F_{2L}F_{1L}}{4}
\left[ -2 \hat{t}^2 D_{12} x \overline{d}(x, Q^2) \right]
\label{int3} \\
\frac{d^2 \sigma_{CC}(e^+_{R}p)}{dx\;dQ^2}
=   \frac{2\pi \alpha^2}{xQ^4}  \frac{F_{2L}F_{1L}}{4}
\left[ -2 \hat{t}^2 D_{12} x d(x, Q^2) \right]
\label{int4} 
\eeq
with 
\beq
D_{12}= \frac{(\hat{s}-m_1^2)(\hat{s}-m_2^2) + m_1 m_2 \Gamma_1 \Gamma_2}{[
(\hat{s}-m_1^2)^2 + m_1^2 \Gamma_1^2] [(\hat{s}-m_2^2)^2 + m_2^2 \Gamma_2^2]}
\eeq
where $m_1$ ($m_2$) is  the mass of $\ptp$ ($\pto$), and the $\Gamma_1$ 
($\Gamma_2$) 
is the corresponding decay width. The numerical calculation shows that the 
cross section due to the above interference term is very small for $\Delta 
m=25 GeV$, say $\sim 1 pb$ for both NC and CC processes.

As for the couplings, we have 
$F_{1L}=0.013, 0.02$ and $0.03$ for $F_{2L}=0.02$ and $\beta=0.6,0.5$, 
and $0.4$ respectively. More specifically, we will consider the following 
three sets of branching ratios:

\begin{quotation}
\noindent
Set-A: $\breda =0.6$, $\bredb =0.5$;\\
Set-B: $\breda =0.5$, $\bredb =0.5$;\\
Set-C: $\breda =0.4$, $\bredb =0.4$.
\end{quotation}
All three sets of branching ratios are allowed by CDF and D0 limit
\cite{cdf210,d0225} respectively. However, if we consider the 
combined Tevatron
limit\cite{kramer972}, the Set-A and Set-B are excluded, but
the Set-C is  still allowed. For completeness, we present the
numerical results corresponding to all three  sets of branching ratios in
Table 7 and Table 8.

Table 7 shows the total contributions to NC cross sections from both the
lighter mixed state $\ptp$ and the heavier mixed state $\pto$,
assuming $m(\ptp)=200 GeV$, $m(\pto)=225 GeV$, $F_{2L}=0.02$ and $F_R=0$.
The $\ptp$ and $\pto$ can provide the required
contribution to NC cross sections. For $Q^2_{min}=15000 GeV^2$, for example,
$\slq^A=0.216 pb$ and $\slq^C=0.157 pb$.

If we change the Case-3 mass spectrum to $m(\ptp)=205 GeV$
and $m(\pto)=230 GeV$,  the TC
leptoquark contribution to NC cross section will generally decrease by about
$25\%$, but still large enough to account for the excess of NC events with
high-$Q^2$.

\begin{table}[htbp]
\begin{center}
\caption{The total integrated NC cross sections from TC 
leptoquarks $\ptp$ and
$\pto$ for different $Q^2$ cuts.
The cross sections $\slq^{A,B,C}$ corresponding to the three 
sets of branching ratios respectively.
The  preliminary 1994-97 ZEUS and H1 $e^+p$ NC data  are included.
All cross sections in $pb$.}
\label{table7}
\vskip 0.5pc
\begin{tabular}{|c|c|c|c||c|c|c|}\hline
$Q^2_{min}$ &\multicolumn{2}{c|}{ H1 + ZEUS }& &
 \multicolumn{3}{c|}{ $\ptp$ and $\pto$ Contributions} \\ \hline
($GeV^2$) &$N_{obs}$& $\sigma_{obs}$&$\sigma_{sm}$
&$\slq^{A}$&$\slq^{B}$& $\slq^{C}$ \\ \hline
2500 &724 $^{(a)}$ &$43.3^{+4.6}_{-3.9}$&45.7&0.322&0.277&0.240 \\
5000&193 + 326  $^{(b)}$& $10.7 \pm 0.7$& 10.6&0.311&0.267&0.222 \\
10000&31 + 50 $^{(b)}$& $1.70 ^{+ 0.23}_{-0.20}$&1.79&0.274&0.237&0.189\\
15000&18+ 18 $^{(b)}$&$0.71 ^{+0.14}_{-0.12}$&0.49&0.216&0.197&0.157 \\
20000&7 + 7 $^{(b)}$&$0.30^{+0.092}_{-0.076}$&0.161&0.183&0.158&0.126  \\
25000&4 + 3 $^{(b)}$&$0.16^{+0.069}_{-0.053}$&0.059&0.140&0.121&0.097  \\
30000&2+ 2 $^{(b)}$&$0.098^{+0.059}_{-0.042}$&0.023&0.096&0.084&0.067  \\
35000&2$^{(c)}$&$0.060^{+0.059}_{-0.037}$&0.0091&0.054&0.047&0.038  \\
40000&1$^{(c)}$&$0.032^{+0.044}_{-0.023}$&0.0036&0.012&0.012&0.010
 \\ \hline
\end{tabular}
\end{center}
\noindent
(a). ZEUS 1994-97 $e^+p$ NC data, ${\cal L} = 33.5 pb^{-1}$;\\
(b). Combined H1 and ZEUS 1994-97 
$e^+p$ NC data, ${\cal L } = 57.2 pb^{-1}$; \\
(c). H1 1994-97 $e^+p$ NC data, ${\cal L } = 23.7 pb^{-1}$.
\end{table}
\vspace{.5cm}

Table 8 shows the total leptoquark contribution to CC cross section from
both $\ptp$ and $\pto$, assuming the Case-3 parameters. 
The TC leptoquark
$\ptp$ and $\pto$ can provide the required contributions to account for both
excesses of NC and CC events with very high $Q^2$ simultaneously. 
For $Q^2_{min}
=15000 GeV^2$, for instance, we have $\slq^A=0.216 pb$ and $\scc^A=0.553 pb$
respectively.  The lighter $\ptp$ again dominates the total CC contribution,
about $75\%$ of $\scc$ comes from the lighter $\ptp$. 
While the size of $\slq^A$ is just what we want  to get,
the CC contribution $\scc^A$ seems to be a bit large when compared with the
corresponding $\Delta \sigma_{CC}$ as shown in Table 8. 
The difference becomes more apparent  for other two sets of branching ratios.

Fig.2 and Fig.3 show the NC and CC  cross sections from  different sources
as a function of $Q^2$ cut,
assuming $m(\ptp)=200 GeV$, $m(\pto)=225 GeV$, $F_{2L}=0.02$, $F_R=0$ and
Set-B branching ratios. The $\sigma_{SM}$ represents the SM contribution,
while the $\sigma_1$ and $\sigma_2$ shows the contribution from $\ptp$ and
$\pto$ respectively. For NC cross section, the $\sigma_{tot}$ now is in 
very good agreement with the combined 1994-97 H1 and ZEUS NC data.
For both NC and CC process, the lighter $\ptp$ dominates: 
about $70\%$ of the total leptoquark cross section
is due to the lighter $\ptp$ for $Q^2_{min} < 35000 GeV^2$.

For $\emp$ collision process, the contributions from both
$\ptp$ and $\pto$ are negligibly  small when 
compared with that to $\epp$ process
 because of the strong suppression by the smallness of parton density of 
sea quarks in a proton.

\begin{table}[htbp]
\begin{center}
\caption{The total leptoquark  contributions to CC cross sections
 with different $Q^2_{min}$ cuts, assuming the Case-3 parameters.
The CC cross sections  $\scc^{A,B,C}$ corresponding to
three  sets of branching ratios respectively.
The ZEUS and H1 preliminary 1994-97 $e^+p$ CC data included.
All cross sections in $pb$.}
\label{table8}
\vskip 0.5pc
\begin{tabular}{|c||c|c|c|c||c|c|c|}\hline
$Q^2_{min}$ &\multicolumn{2}{c|}{ H1 + ZEUS }& & &
 \multicolumn{3}{c|}{ $\ptp$ and $\pto$ Contributions} \\ \hline
($GeV^2$) &$N_{obs}$& $N_{expect}$& $\sigma_{SM}$
&$\triangle \sigma_{CC}$&$\scc^{A}$& $\scc^{B}$& $\scc^C$ \\ \hline
1000  &455&$419 \pm 36$    &16.47 &$1.4  \pm 1.4$ &0.694&0.824&0.989 \\
2500  &61 & $56.3 \pm 9.4$ &7.32 &$0.61  \pm 1.2$ &0.692&0.821&0.986 \\
5000  &43 &$34.7 \pm 6.9$  &2.54 &$0.61  \pm 0.51$ &0.679& 0.806&0.967\\
10000 &28 & $17.7 \pm 4.3$ &0.493 &$0.287\pm 0.12$&0.628&0.744&0.893\\
15000 &11 &$4.9 \pm 1.7$   &0.127 &$0.158\pm 0.04$&0.553&0.655&0.787\\
20000 &5  & $1.7 \pm 0.7$  &0.037 &$0.072\pm 0.015$&0.466&0.550&0.660 \\
30000 &1  &$0.034 ^{+0.038}_{-0.018}$&0.004&$0.114^{+0.004}_{-0.002}$
&0.266&0.310&0.372\\\hline
\end{tabular}
\end{center}
\end{table}

\section*{4. Conclusion and discussions}     

In this paper we try to pursue a consistent TC 
 leptoquark interpretation for both 
excesses of NC and CC events with very high $Q^2$ observed by H1 and ZEUS 
collaborations.  

In the framework of Technicolor, the F=0 scalar leptoquark 
$\ptp$ and $\pto$ are produced naturally by mixing of 
charge $2/3$ color-triplet pseudo-Goldstone bosons $\punb$ and $\pdeb$, 
and therefore they can contribute to both NC and CC processes simultaneously 
through  their decays to $e^+d$ and $\overline{\nu}u$ final states.

The size of the TC leptoquark contributions  strongly depends on the mass
spectrum of leptoquarks and the value of effective Yukawa couplings. If the 
Yukawa couplings of TC leptoquarks to first generation lepton-quark pairs was
proportional to light fermion mass, say 
$\lambda \approx (m_u + m_e)/F_\pi$, the corresponding TC leptoquark 
contribution to NC and CC processes would be too small to account for 
the observed excesses at HERA.

If we assume, following the general BRW leptoquark scenario 
\cite{buchmuller87}, that the effective 
Yukawa couplings of TC leptoquarks to light lepton-quark pairs are also the 
dimensionless parameters being constrained by the known experiments, the TC 
leptoquarks with mixing 
can indeed provide the required extra contributions to both NC and
CC $e^+p$ processes. For $\emp$ collision process, on the contrary,  
the contributions  from F=0 scalar TC leptoquarks
are indeed very small and can be neglected safely.

The charge $5/3$ TC leptoquark $P_3^1$ can provide an adequate contribution 
to NC process, but it could not contribute to CC process at HERA.

The charge $2/3$ TC leptoquarks $\ptp$ and $\pto$ 
can contribute to both NC and CC processes simultaneously. 
But they could not be light at the same time because of the stringent limits
from Tevatron experiments. 
Using the parameters allowed by all known  experiments, 
one single F=0 scalar TC leptoquark $\ptp$ with $m(\ptp)=200 GeV$ 
can provide the required contributions 
to account for the observed excesses of both NC and CC events with 
very high $Q^2$ simultaneously, 
as shown in Tables (4,5,6). For $Q^2_{min} =15000 GeV^2$, we have $\slq^a=
0.217 pb$ and $\scc^a = 0.306 pb$,  while  about $0.2 pb$ extra NC and CC 
contributions 
are  required to explain the observed NC and CC excess respectively.

Inspired by the apparent splitting of the average mass for H1 
and ZEUS NC events, we suppose  that: there may exist two
scalar TC leptoquarks $\ptp$ and $\pto$, they are relatively light with
the masses of $m(\ptp)=200 GeV$ and $m(\pto)=225 GeV$. Using  the parameters
 allowed by current experimental constraints, we
found that $\ptp$ and $\pto$ can provide the required contributions 
to  NC and CC $\epp$ processes
simultaneously. For $Q^2_{min} =15000 GeV^2$,
we have $\slq^A=
0.216 pb$ and $\scc^A = 0.553 pb$. The size of $\slq^A$ is just what we
need, but the CC cross sections $\scc^{A,B,C}$ seem to be larger
than that required. For $ m(\ptp)=205 GeV$ and $m(\pto)=230 GeV$, the total
NC and CC contributions  will in general decrease  by about $25\%$.

\vspace{.5cm}
\noindent {\bf ACKNOWLEDGMENT}

Z. Xiao would like to thank
R.G.Roberts, H.Dreiner for help in computing and discussions on the 
current topic.  Many thanks to F.E.Close, 
G.Moley, H.M. Chan and R.J.N. Phillips for their encouragement and help. 
It is a pleasure for me to thank the Rutherford-Appleton Laboratory for 
hospitality and support. 
This work is  supported  by the National Natural Science 
Foundation of China and  by the Sino-British Friendship Scholarship Scheme.


\newpage
\begin{center}
{\bf Figure Captions}
\end{center}
\begin{description}

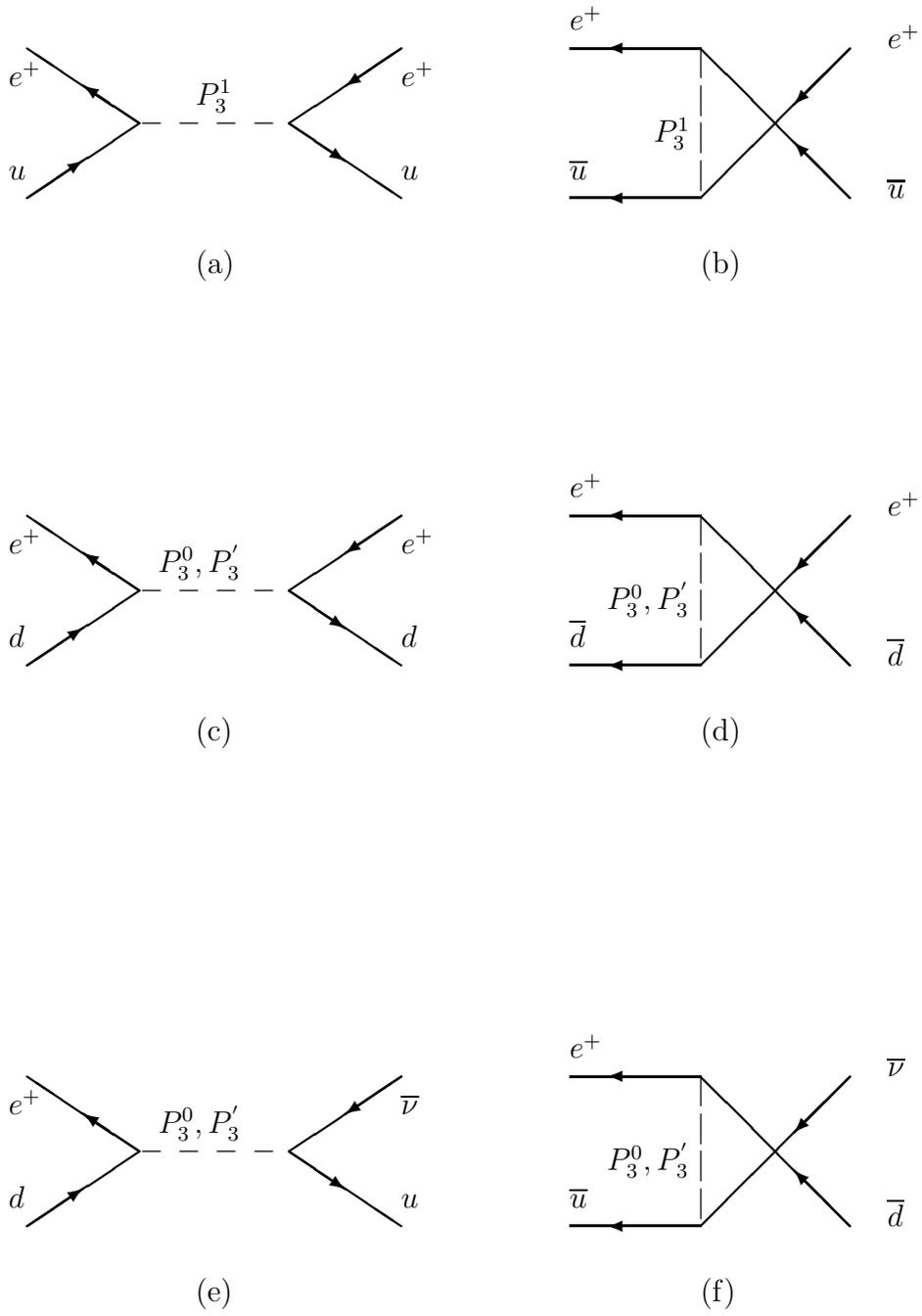
\item[Fig.1:] The Feynman diagrams for the production and decay of F=0 
scalar TC leptoquarks through s-channel and u-channel.

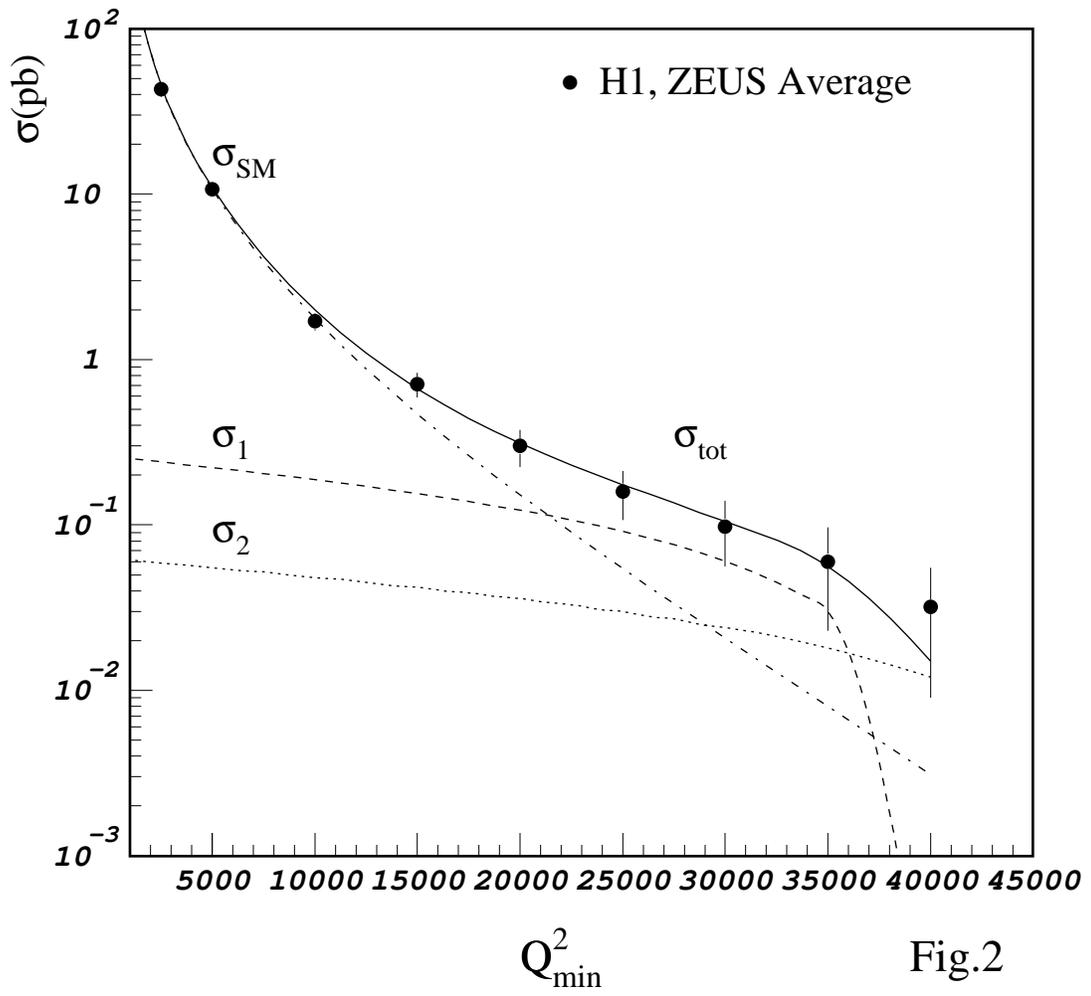
\item[Fig.2:] The neutral current cross sections from the SM $(\gamma, Z)$ 
gauge bosons and the F=0
scalar TC leptoquarks $\ptp$ and $\pto$ for $Q^2 > Q^2_{min}$, assuming 
the Case-3 parameters. The $\sigma_1$ and $\sigma_2$ shows 
the contribution from the $\ptp$ and 
$\pto$, respectively. $\sigma_{tot}=\sigma_{SM} + \sigma_1 + \sigma_2$ 
is the total NC cross section.  

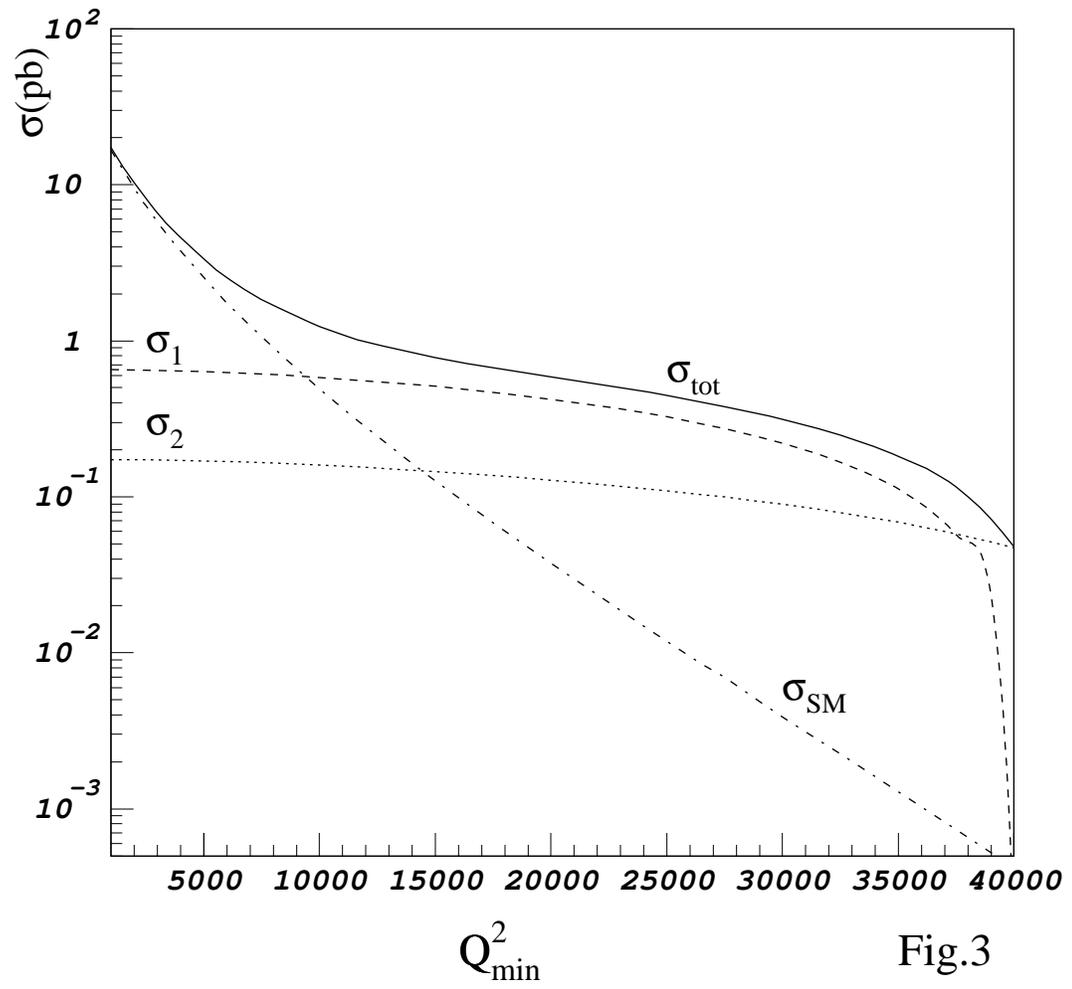
\item[Fig.3:] The charged  current cross sections from the SM  W gauge boson 
and the F=0
scalar TC leptoquarks $\ptp$ and $\pto$ for $Q^2 > Q^2_{min}$, assuming
the Case-3 parameters. The $\sigma_1$ and $\sigma_2$ shows
the contribution from the $\ptp$ and
$\pto$, respectively. The $\sigma_{tot}$ is the total CC cross section.

\end{description}

\newpage

\begin{picture}(30,0)
{\thicklines
\setlength{\unitlength}{0.1in}

\put(12,-7){\vector(-3,2){3}}
\put(12,-7){\line(-3,2){6}}
\put(6,-11){\vector(3,2){3}}
\put(12,-7){\line(-3,-2){6}}
\multiput(12,-7.4)(2,0){4}{$-$}
\put(26,-3){\vector(-3,-2){3}}
\put(20,-7){\line(3,2){6}}
\put(20,-7){\vector(3,-2){3}}
\put(20,-7){\line(3,-2){6}}
\put(5,-5){$e^+$}  
\put(5,-10){$u$}
\put(26,-5){$e^+$}
\put(26,-10){$u$}
\put(15,-6){$P_3^1$}
\put(15,-15){(a)}

\put(42,-3){\vector(-1,0){5}}
\put(35,-3){\line(1,0){7}}
\put(42,-11){\vector(-1,0){5}}
\put(35,-11){\line(1,0){7}}
\multiput(41.8,-4.2)(0,-2){4}{$\vert$}
\put(50,-3){\vector(-1,-1){3}}
\put(42,-11){\line(1,1){8}}
\put(50,-11){\vector(-1,1){3}}
\put(42,-3){\line(1,-1){8}}
\put(35,-2){$e^+$}
\put(35,-10){$\overline{u}$}
\put(52,-3){$e^+$}
\put(52,-11){$\overline{u}$}
\put(39.5,-8){$P_3^1$}
\put(42,-15){(b)}

\put(12,-32){\vector(-3,2){3}}
\put(12,-32){\line(-3,2){6}}
\put(6,-36){\vector(3,2){3}}
\put(12,-32){\line(-3,-2){6}}
\multiput(12,-32.4)(2,0){4}{$-$}
\put(26,-28){\vector(-3,-2){3}}
\put(20,-32){\line(3,2){6}}
\put(20,-32){\vector(3,-2){3}}
\put(20,-32){\line(3,-2){6}}
\put(5,-30){$e^+$}
\put(5,-35){$d$}
\put(26,-30){$e^+ $}
\put(26,-35){$d$}
\put(13,-31){$P_3^0, P_3^{'}$}
\put(15,-40){(c)}
  
\put(42,-28){\vector(-1,0){5}}
\put(35,-28){\line(1,0){7}}
\put(42,-36){\vector(-1,0){5}}
\put(35,-36){\line(1,0){7}}
\multiput(41.8,-29.2)(0,-2){4}{$\vert$}
\put(50,-28){\vector(-1,-1){3}}
\put(42,-36){\line(1,1){8}}
\put(50,-36){\vector(-1,1){3}}
\put(42,-28){\line(1,-1){8}}
\put(35,-27){$e^+$}
\put(35,-35){$\overline{d}$}
\put(52,-28){$e^+$}
\put(52,-36){$\overline{d}$}
\put(37,-33){$P_3^0, P_3^{'}$}
\put(42,-40){(d)}

\put(12,-62){\vector(-3,2){3}}
\put(12,-62){\line(-3,2){6}}
\put(6,-66){\vector(3,2){3}}
\put(12,-62){\line(-3,-2){6}}
\multiput(12,-62.4)(2,0){4}{$-$}
\put(26,-58){\vector(-3,-2){3}}
\put(20,-62){\line(3,2){6}}
\put(20,-62){\vector(3,-2){3}}
\put(20,-62){\line(3,-2){6}}
\put(5,-60){$e^+$}
\put(5,-65){$d$}
\put(26,-60){$\overline{\nu}$}
\put(26,-65){$u$}
\put(13,-61){$P_3^0, P_3^{'}$}
\put(15,-70){(e)}

\put(42,-58){\vector(-1,0){5}}
\put(35,-58){\line(1,0){7}}
\put(42,-66){\vector(-1,0){5}}
\put(35,-66){\line(1,0){7}}
\multiput(41.8,-59.2)(0,-2){4}{$\vert$}
\put(50,-58){\vector(-1,-1){3}}
\put(42,-66){\line(1,1){8}}
\put(50,-66){\vector(-1,1){3}}
\put(42,-58){\line(1,-1){8}}
\put(35,-57){$e^+$}
\put(35,-65){$\overline{u}$}
\put(52,-58){$\overline{\nu}$}
\put(52,-66){$\overline{d}$}
\put(37,-63){$P_3^0, P_3^{'}$}
\put(42,-70){(f)}
\put(28,-80){Fig.1}
 
}

\end{picture}

\end{document}